\documentclass[12pt]{article}
\usepackage{etex}
\usepackage{amsfonts}
\usepackage{latexsym}
\usepackage{natbib}
\usepackage{amsmath, pst-pdf, pst-all, amsthm, amssymb, latexsym}
\usepackage{comment}
\usepackage{listings}
\usepackage{pstricks,pstricks-add}
\usepackage{booktabs}
\usepackage{rotating}
\usepackage{multirow, tabularx}
\usepackage{lscape}
\usepackage{abstract}
\usepackage{lipsum} 
\usepackage{ctable}
\def\sym#1{\ifmmode^{#1}\else\(^{#1}\)\fi}
\usepackage{caption}
\usepackage{subfloat}
\usepackage{titlesec}
\usepackage{scalerel}
\newcommand\vfrac[2]{\ThisStyle{%
  \setbox0=\hbox{$\SavedStyle#1#2$}%
  \setbox2=\hbox{$\SavedStyle X$}%
  \ifdim\ht0>\ht2\setlength{\ht0}{\ht2}\fi%
  #1\mathord{\stretchto{\raisebox{2.3\LMpt}{$\SavedStyle/$}}{\ht0}}#2}}

\usepackage{multirow,array}
    \let\quoteOLD\quote
    \def\quote{\quoteOLD\small}
    
   \usepackage{blindtext}
\renewenvironment{quote}
  {\small\list{}{\rightmargin=0.85cm \leftmargin=0.9cm}%
   \item\relax}
  {\endlist}



\makeatother
\usepackage{indentfirst}

\usepackage{enumerate}

\usepackage{hyperref}
\hypersetup{backref,  
pdfpagemode=FullScreen,  
colorlinks=true, citecolor=blue!50!black}
\usepackage{graphicx}

\usepackage{sgame}

\usepackage{pst-all}
\usepackage{fancyhdr}

\usepackage{enumitem}

\newtheorem*{prop1}{Proposition 1} 
  \newtheorem*{assumption1}{Assumption 1} 
    \newtheorem*{assumption2}{Assumption 2} 
    \newtheorem*{assumption3}{Assumption 3}
           
  \titlespacing\section{0pt}{10pt plus 4pt minus 2pt}{8.5pt plus 2pt minus 2pt}
\titlespacing\subsection{0pt}{10pt plus 4pt minus 2pt}{9pt plus 2pt minus 2pt}
\titlespacing\subsubsection{0pt}{10pt plus 4pt minus 2pt}{8.5pt plus 2pt minus 2pt}

\begin{document}

\title{Conflict externalization and the quest for peace: Theory and case evidence from Colombia\thanks{I am grateful to  Raphael Godefroy, David Karp, Alessandro Riboni and two anonymous referees for helpful comments and suggestions.  All errors or omissions are mine.}}
\author{Hector Galindo-Silva\thanks{%
Department of Economics, Pontificia Universidad Javeriana, E-mail: galindoh@javeriana.edu.co
} \\
Pontificia Universidad Javeriana\\ 
}
\date{First Draft: June 2014 \\This Draft: August 2020}
 \maketitle
\begin{abstract}
I study the relationship between the likelihood of a violent domestic conflict and the risk that such a conflict ``externalizes'' (i.e. spreads to another country by creating an international dispute).  I consider a situation in which a domestic conflict between a government and a rebel group has the potential to externalize.  I show that the risk of externalization increases the likelihood of a peaceful outcome, but only if the government is sufficiently powerful relative to the rebels, the risk of externalization is sufficiently high, and the foreign actor who can intervene in the domestic conflict is sufficiently uninterested in material costs and benefits.  I show how this model helps to understand the recent and successful peace process between the Colombian government and the country's most powerful rebel group, the Revolutionary Armed Forces of Colombia (FARC). 

\emph{Journal of Economic Literature} Classification Numbers: H72, D72

\bigskip
\noindent \emph{Keywords:} Conflict, externalization, peace talks

\end{abstract}

\newpage
\section{Introduction}

The existing theoretical literature on civil wars often assumes that a group's decision of whether to fight  depends exclusively on the domestic context, i.e. on other \emph{domestic} groups' decisions.\footnote{See \cite{BlattmanMiguel2010} and \cite{JacksonMorelli2011handbook} for reviews of this literature.}  However, an increasing empirical literature on conflict is demonstrating that the regional context also plays an important role in domestic conflicts.\footnote{See \cite{MasonWeingartenFett1999}, \cite{HegreSambanis2006}, \cite{Gleditsch2007}, \cite{BalchLindsayDylanEnterlineJoyce2008} and \cite{CunninghamGleditschSalehyan2009, CunninghamGleditschSalehyan2011} for evidence showing that the regional context (e.g. a conflict in a neighboring country, a highly autocratic region, trans-boundary ethnic groups, or direct intervention of external parties) matters for the onset, incidence and duration of civil wars.} Researchers are well aware of the regional dimensions of conflicts in Myanmar, Nicaragua, Kosovo, Sudan, Lebanon and Iraq, for example.\footnote{See \cite{South2008} for the conflict in Myanmar; \cite{GleditschBeardsley2004} for the Nicaragua-Contras conflict; \cite{Crawford2001} and \cite{Kuperman2008} for Kosovo; \cite{AliElbadawiElBatahani2005} for Sudan; \cite{BouckaertHoury2007} for Lebanon; and \cite{Gunter2008} and \cite{MorelliPischedda2013} for the Iraqi-Kurdish conflict.}

In this paper, I develop a simple model of conflict externalization, and provide new case-study evidence from Colombia. The objective of the model is to formalize an existing theory, and to propose a new mechanism through which the possibility that an external party intervenes in an internal conflict could affect the outcome of the conflict. In addition to demonstrating the applicability of the model, the Colombian case provides new evidence on this topic.  

In the model, a government and a group of rebels simultaneously choose whether to attack each other. Attacking is costly, but can also decrease the opponent's military resources, which increases the aggressor's probability of victory. Crucially, the use of violence might cause a third (foreign) actor to join in the conflict. When this happens, I say that the domestic conflict ``externalizes.'' 
 
This externalization changes the power dynamic between the two domestic actors. I assume that only the government's use of violence can trigger such an externalization. For example, consider a group of rebels strategically located along a porous border, where the neighboring country shares an ideology or ethnicity with the rebels. Given the relationship between the rebels and the neighboring country's government, an aggression might be viewed as a violation of the neighboring country's sovereignty and could motivate a military response, starting a conflict spiral that might lead to an international war.\footnote{As I will argue later, this situation matches the recent dynamic of the Colombian internal conflict. It is also consistent with Myanmar-Thailand border clashes prompted by Myanmar pursuit of Karen National Liberation Army rebels across the border into Thailand; see \cite{South2008, South2012}.} I focus on how the threat of such an external conflict affects the likelihood of peace between domestic actors.

This paper's first main contribution is to show that the risk of externalization increases the likelihood of peace, but that this only happens if a government is sufficiently powerful relative to rebels, if the risk of externalization is sufficiently high, and if a foreign actor is sufficiently uninterested in material costs and benefits. 

This paper's second main contribution is new case-study evidence from Colombia. Colombia has suffered one of the world's longest-running internal conflicts. Although many armed groups have participated in the conflict, the left-wing Revolutionary Armed Forces of Colombia (FARC) was always the largest non-government actor. In September 2012, the Colombian government announced the start of new peace negotiations following an extensive military campaign that severely damaged the FARC. This announcement surprised many analysts and national leaders, who expected a few more years of war, ending with a government victory. In July 2016, the Colombian government and the FARC signed a historic peace deal, which earned then-Colombian President Juan Manuel Santos the 2016 Nobel Peace Prize.

Despite its relevance, to the best of my knowledge no rationalist explanation has been proposed to account for both the onset and success of this Colombia's historic peace process with the FARC. I show that the risk of externalization of the Colombian conflict (to Venezuela) was at the root of the peace negotiations, creating what has been called a situation that was ``ripe for resolution'' \cite{Zartman2000}.

The theory proposed in this paper relates to the literature studying the interdependence of intra- and inter-group conflicts  \cite[see for instance][]{BaikLee2000, SteinRapoport2004, Hausken2005, Munster2007, MunsterStaal2011, ChoiChowdhuryKim2016}. Using models of group contest, these studies examine conflicts within or between groups that occur either sequentially or simultaneously. This paper uses a contest model with sequential actions \cite[as in][]{BaikLee2000, SteinRapoport2004, MunsterStaal2011}, with heterogeneity within groups  \cite[as in][]{ChoiChowdhuryKim2016}. However, those studies focus on scenarios in which all members of a domestic group have a common interest in fighting a foreign enemy, whereas this paper focuses on a scenario with two domestic groups, where one group is closely aligned with a foreign group and the other is not.

This paper is also related to the few but increasing theoretical studies on third-party interventions \cite[see][]{Fearon1998, CarmentRowlands1998, Werner2000, Crawford2003, CarmentRowlands2006, AmegashieKutsoati2007, Munster2007, MunsterStaal2011, KyddStraus2013}.  Specifically, it is consistent with what this literature calls the ``deterrence'' hypothesis: the idea that an external party can play a crucial role in the outcome of an internal conflict by deterring one of the domestic parties from making a decision that harms its opponent \citep[see][]{Fearon1998, CarmentRowlands1998, Werner2000, Crawford2003}. In my model, peace is possible because the risk of externalization deters the government from attacking the rebels, given that it could prompt externalization that could strengthen the rebels.\footnote{Although the argument does not account for ``moral hazard'' --- that an external intervention biased in favor of one party might make this party more belligerent  --- the model in this paper can be easily extended to include this possibility, which does not change the key findings.  However, this phenomenon does not seem to play an important role in the Colombian conflict, and despite its simplicity, the model is able to provide non-trivial and new and empirical predictions.} As in \cite{KyddStraus2013}, I find that whether the potential for an external intervention makes war less likely depends on whether power is balanced, but unlike them, I find that peace is more likely when the government is stronger, even if an intervention strengthens the rebel group.\footnote{This paper also shares similarities with \cite{BakshiDasgupta2019},  who study how the balance of power between groups within a country affects group conflict in another country. Like \citeauthor{BakshiDasgupta2019}, this paper also studies the relationship between cross-border spillovers and conflict. However, unlike \cite{BakshiDasgupta2019}, the results in this paper crucially depend on the possibility of interstate conflict.} 

The outline of the paper is as follows. Section \ref{model} presents the model. Section \ref{caseevidence} presents and analyzes the case study based on the model. Section \ref{conclusion} concludes.

\section{Model}\label{model}

In this section, I develop a simple model that illustrates how the risk of a domestic conflict externalizing can affect the outcome of the conflict. 

Suppose that a government  faces a rebel group. The government's main goal is to defeat the rebels. The rebels are not strong enough to defeat the government, but can harm the government and thereby avoid being defeated. Both parties simultaneously choose whether to attack the other or pursue peace. If either party attacks, the other party suffers a loss $L$ in military resources, and both parties lose an amount of wealth, $C$.\footnote{The assumption that $C$ and $L$ are the same for both groups seem restrictive. However, the results are robust to potential differences, within certain constraints.}  Importantly, if the government attacks the rebels, then a foreign government $F$ can intervene. This intervention is assumed to favor the rebel group,\footnote{The assumption that an external intervention can occur only if the government attacks the rebels is crucial for the main result (that, under certain conditions, peace is selected as the unique equilibrium). If an external intervention could also occur when the rebels attack the government, then conflict would always be an equilibrium, and the conditions for an additional peaceful equilibrium would be harder to achieve. Thus, this alternative scenario would imply weaker results. However, the mechanism would be the same. The details for this alternative scenario are available upon request.}  and always implies that the  government is unable to defeat the rebel group.\footnote{For an intuition of the types of situations I try to model, consider a domestic conflict where some rebels are located close to a border shared with $F$. Given the proximity of the rebels to $F$, and the ability of the rebels to move between the two countries (because of either porous borders or sympathy from $F$), an offensive action by the government in the rebels's controlled territory might harm the citizens of $F$ or be interpreted by $F$ as a violation of its sovereignty. This implies a strong reaction by $F$ to try and ensure that the government is unable to defeat the rebels. This intervention can be interpreted either as a direct attack on the government, or as military aid to the rebels.  In Section \ref{caseevidence}, I argue that this scenario is consistent with what occurred prior to the peace process between the Colombian government and the FARC.} 

The model focuses on the role of the government's initial resources in determining the outcome of the game. I use $G$ to denote these resources. I assume that $G\in(L,\overline{G})$.\footnote{The fact that $G> L$ implies that the harm caused by the government's violence is limited by the government's resources. The condition $\overline{G}>G$ defines an upper bound for $G$.} The probability of the government and the rebels winning, for all outcomes in which $F$ does not intervene, depends on $G$. To model these probabilities, I first define a function $Z$ such that $Z(x)=0$ for all $x\leq 0$, so that $Z' = 0$ for all $x < 0$; $Z'>0$ and $Z''\leq 0$ for all $x\in(0,\overline{G})$; and $Z(x)=1$ for all $x\geq \overline{G}$. This function will be key in the analysis, and can interpreted as the government's probability of defeating the rebels if the group with the most resources wins,\footnote{This means that  if the resources of the groups $A$ and $B$ are $R_A$ and $R_B$, respectively, then the probability of victory for group $A$ is 1 if $R_A>R_B$, 0 if $R_A<R_B$ and $1/2$ if $R_A=R_B$. This form of modeling the winning probabilities is known as an ``all-pay auction.'' It has been used to study contests in which group members exert effort that translate into ``group effort'' \cite[see for instance][]{BaikKimNa2001, BarbieriMaluegTopolyan2014, ChowdhuryLeeTopolyan2016, ChowdhuryTopolyan2016a, ChowdhuryTopolyan2016b}. Under this interpretation, the scenarios in which one group defends against another group's attack \cite[as in][]{ChowdhuryTopolyan2016a}, and where the aggregation technology is asymmetric among the contesting groups \cite[as in][]{ChowdhuryTopolyan2016b} are particularly important.}  and if the resources of the rebels are random.\footnote{To see this, define $R$ as the resources of the rebels, and assume that $R$ is a random variable on $\mathbb{R}$ with a cumulative distribution function $Z$ such that  $Z(x)=0$ for all $x\leq 0$, $Z'>0$,$Z''\leq 0$ for all $x\in(0,\overline{G})$, and $Z(x)=1$ for all $x\geq \overline{G}$. Thus, if the winning probabilities are defined as an ``all-pay auction,'' we have that the government's probability of victory is $Pr[G>R]=Z(G)$ if both groups attack. The other winning probabilities can be defined analogously. The randomness of $R$ is consistent with the rebels being somewhat informal. This may be particularly true if the rebels are located along the borders with $F$, if these borders are porous, and if there is uncertainty about how effectively $F$ surveils its border (as the level of arms or troops flowing in from $F$ may be unknown).} 

Specifically, let $Z(G)$ be the government's probability of defeating the rebels when both parties attack and $F$ does not intervene. When the rebels attack but the government chooses peace, the government loses $L$ in military resources, so its probability of winning is $Z(G-L)$. The winning probabilities are defined analogously when the government attacks but neither the rebels nor the foreign country attack.\footnote{As previously mentioned, when $F$ intervenes, the government's probability of defeating the rebels is $0$.} When no group attacks, peace occurs; in this scenario, the parties compete in an election in which each party's probability of winning  depends on the relative initial resources.\footnote{This can be interpreted as an scenario in which the political parties that represent the interests of the groups compete for seats in the national assembly under a proportional representation system. It is also consistent with a number of seats in the national assembly being given to the rebels under the terms of a peace deal that guarantees proportional representation for the rebels. These scenarios are consistent with what occurred after the 2016 peace accord between the Colombian government and the FARC.} 

Finally, $F$'s decision of whether to intervene depends negatively on the resources of the government: the greater these resources, the less likely it is that $F$ will intervene because $F$'s potential losses are greater. However, and importantly,  $F$ may not be purely interested in material costs and benefits. $F$ may have ideological or religious motivations, which, insofar as they involve agents that can be difficult to negotiate with, can be viewed as being not materially based \cite[in this respect, see][]{JacksonMorelli2011handbook}.\footnote{As will be discussed in Section \ref{caseevidence}, ideology is crucial to understanding Venezuela's motivation to intervene in the Colombian conflict. In addition to \cite{JacksonMorelli2011handbook}, see \cite{Maynard2019},  who discusses the role of ideology in armed conflicts, and \cite{Owen2010}, who provides examples in international politics.}  Specifically, $F$'s goals are purely material with probability $1-\phi$, in which case its decision to intervene depends on $G$. $F$'s goals are not materially based with probability $\phi$,  in which case the probability of an intervention is assumed to be exogenous and normalized to one.\footnote{This normalization is without loss of generality; what matters is that the probability of an intervention when $F$'s goals are not materially based is exogenous.}  

\subsection*{Timing}

  \begin{enumerate}\itemsep0em 
\item[(1)] The government and the rebels simultaneously decide whether to attack each other. 
\item[(2)] If either party decides to attack, a violent conflict occurs. If neither party attacks, peace occurs, and the parties compete in an election in which each party's probability of winning depends on its relative initial resources.
\item[(3)] If there is a violent conflict, both parties face exogenous costs $C>0$. In addition, the military resources of a party decrease by $L$ if it is attacked.
\item[(4)] $F$ observes whether the government attacked. If the government attacked, $F$ intervenes with probability $\phi$, and makes a determination on whether or not to intervene based on G with probability $(1-\phi)$.  If $F$ intervenes, the government cannot defeat the rebels. 
\item[(5)] The violent conflict ends. Each group receives a payoff according to the outcome, which is described in the next subsection. 
\end{enumerate}

\subsection*{Payoffs}

The preferences of the domestic groups are characterized by a Bernoulli utility function, where each party gets a payoff of 1 if it wins, and 0 if it loses. The payoff function of the government is denoted by $\pi$, and the payoff function of the rebels is denoted by $\rho$. Note that these functions depend on what the domestic groups simultaneously decide (either attack $(a)$ or peace $(p)$), as well as on the foreign government's decision about whether to intervene $(a)$ or not $(p)$. 

Since the government and rebels do not know whether $F$ will intervene when deciding on their military actions, I use $\Phi$ to denote each group's expectation that such an intervention will occur (conditional on an attack by the government). Thus, given an attack by the rebels, the government's expected payoff from attacking is $\pi(a,a)=(1-\Phi) Z(G)-C$.\footnote{Note that, with some abuse of notation, I have defined $\pi(a,a)\equiv E_x[\pi(a,a,x)]$, where $x\in\{a,p\}$ is $F$'s choice, and $E_x$ is the government's expectation about $x$.} The expected payoffs for all cases, after rearranging and omitting some constants, are given in Table \ref{payoffs}.

\vspace{-0.1cm}
 {\renewcommand{\arraystretch}{0.9} 
  \begin{table}[h]
   \begin{center}\small
    \setlength{\extrarowheight}{1.6pt}
    \begin{tabular}{c c | p{6cm} | p{6cm} |}
      & \multicolumn{1}{c}{} & \multicolumn{2}{c}{$Rebels$}\\
      & \multicolumn{1}{c}{} & \multicolumn{1}{c}{$(a)$}  & \multicolumn{1}{c}{$(p)$} \\\cline{3-4}
      \multirow{4}*{$Govt.$}  & \multirow{2}*{$(a)$}  & $\pi=(1-\Phi) Z(G)-C$ & $\pi=(1-\Phi) Z(G+L)-C$ \\
      & & \multicolumn{1}{|r|}{$\rho=-(1-\Phi) Z(G)-C$} & \multicolumn{1}{|r|}{$\rho=-(1-\Phi) Z(G+L)-C$} \\\cline{3-4}
      &  \multirow{2}*{$(p)$}   & $\pi=Z(G-L)-C$ & $\pi=Z(G)$ \\
      & &  \multicolumn{1}{|r|}{$\rho=-Z(G-L)-C$} & \multicolumn{1}{|r|}{$\rho=-Z(G)$} \\
      \cline{3-4}
    \end{tabular}
    \caption{Expected payoffs for government and rebels}\label{payoffs}
    \end{center}  \normalsize
  \end{table}
  }
\vspace{-0.6cm}

I model $F$'s decision about whether to intervene in a very simple and stylized way.  First, I assume that a foreign intervention is less likely when the government is stronger. Then, I define $W(G)$ as the probability of an intervention from the perspective of the government and rebels when $F$ is interested in material benefits, and where $W$ is a function such that $W(x)=0$ for all $x\geq A>\overline{G}$, $W(x)=1$ for all $x\leq 0$, and for all  $x\in(0,A)$, $W'<0$ and $W''\leq 0$.\footnote{A very simple micro-foundation for $W$ is the following. First, define  $F$'s payoff as equal to $B-G$  if $F$ intervenes, and $0$ if it does not intervene, where $B$ represents any material benefit derived from the intervention. Then, assume that $B$ is unknown to the government and the rebels and that, from the point of view of these parties, it has a cumulative distribution function $V$ defined on $\mathbb{R}$ such that  $V(x)=1$ for all $x\geq A>\overline{G}$, $V(x)=0$ for all $x\leq 0$, and for all  $x\in(0,A)$, $V'>0$ and $V''\geq 0$. Finally, set $W(x)\equiv 1-V(x)$.}

\subsection*{Additional assumptions}

\begin{assumption1}
\label{assumption1}
$C>Z(L)$.
\end{assumption1}
\vspace{-0.03cm}

Assumption 1  guarantees that the cost of using violence is high enough that a peace agreement is always possible. 

\begin{assumption2}
\label{assumption2}
 Let sup $\big\{\frac{Z'(G)}{W'(G)}|G\in(L,\overline{G})\big\}\equiv k$. Then $W(\overline{G})k< -1$.
\end{assumption2}

Assumption 2 is a technical condition, and allows for a simpler characterization of the government's best response to a rebel attack.\footnote{This assumption can be relaxed at the cost of much less general functions $Z$ and $W$. Two functions $Z$ and $W$ that satisfy Assumption 2 are:  $Z(x)=\vfrac{x}{\overline{G}}$ if $x\in[0,\overline{G}]$, $Z(x)=0$ if $x<0$, $Z(x)=1$ if $x>\overline{G}$, and $W(x)=(1-\vfrac{x}{A})$ for $x\in[0,A]$ and with $A>2\overline{G}$. More generally, when  $Z(x)=(\vfrac{x}{\overline{G}})^\beta$ with $\beta\in(0,1]$, and $W(x)=(1-\vfrac{x}{A})^\gamma$ with $\gamma\in(0,1]$, then it is easy to see that Assumption 2 holds if $(\frac{\beta}{\gamma})(\frac{A-\overline{G}}{\overline{G}})>1$.  I thank a referee for suggesting this Assumption and example.} 

\begin{assumption3}
\label{assumption3}
$(1-W(\overline{G})) > Z(\overline{G}-L)$.
\end{assumption3}
\vspace{-0.08cm}

Assumption 3 guarantees that given an attack by the rebels, the government's best response function dictates that the government should also attack, even if the probability of a foreign intervention is small.

\subsection{Equilibrium}

I will now characterize the equilibrium of the game. I focus on the intuition for the main equilibrium decisions, and leave the formal derivation to the Appendix. 

First, note that from the point of view of both the government and rebels, a foreign intervention will occur with probability  $\Phi=\phi +(1-\phi)W(G)$ (conditional on the government attacking). Replacing this expression in the payoff functions of the government and the rebels, and computing each group's best response function, it is possible to establish a set of non-trivial conditions under which a conflict occurs in equilibrium. These conditions focus on the exogenous component of the risk of a foreign intervention (i.e. $\phi$), and on the government's resources (i.e. $G$). They constitute the main result of this section, which is summarized in the following proposition:

\begin{prop1}
\label{prop1}
Consider the above-described game. Let $ \overline{\phi}\equiv 1- \frac{Z(\overline{G}-L)}{(1-W(\overline{G}))}$. Given Assumptions 1-3, the following hold:
\vspace{-0.1cm}
\begin{itemize}\itemsep0em 
  \item[(i)] Given any $\phi \in [0,1]$ and any $G\in(L,\overline{G})$, $(p, p)$ must constitute a Nash equilibrium.
 \item[(ii)] Whenever $\phi \leq \overline{\phi}$, $(a, a)$ constitutes a Nash equilibrium for all $G\in(L,\overline{G})$.
 \item[(iii)] Whenever $1 > \phi > \overline{\phi}$, there exists $\hat{G}(\phi)\in(L,\overline{G})$, with $\hat{G}'(\phi) < 0$, such that $(a, a)$ constitutes a Nash equilibrium if and only if $G \leq \hat{G}(\phi)$.
 \item[(iv)] Whenever $\phi = 1$, $(p, p)$ constitutes the unique Nash equilibrium.
 \item[(v)] Neither $(a, p)$ nor $(p, a)$ can constitute a Nash equilibrium.
\end{itemize}
\end{prop1}
\vspace{-0.4cm}
\begin{proof}See Appendix. 
\end{proof}
 
 To understand the intuition behind Proposition 1, first note that if the risk of an external intervention is low, a well-resourced government can easily defeat the rebels. Such a government will have a strong incentive to attack the rebels, so a violent conflict will constitute a Nash equilibrium. In addition, note that this scenario is consistent with the assumption that the risk of an external intervention decreases with the strength of the government. 

However, and importantly, recall that the risk of an external intervention only partially depends on the government's resources. How does this affect the mechanism described in the last paragraph? Crucially, it allows for a non-trivial and new mechanism through which peace can prevail (i.e. in which $(p,p)$ is selected as the unique Nash equilibrium). 
 
This mechanism is based on the idea that when the risk of an external intervention does not entirely depend on the material resources of the groups in conflict, a well-resourced government may have  \emph{less} incentive to attack the rebels. Why? Because this government is better able to tolerate a rebel attack, and by doing so, it can decrease the risk of an external intervention. Importantly, this may happen even if a well-resourced government faces a relatively low probability of an external intervention. For a well-resourced government, the pro-peace effects of resources (a greater capacity to tolerate a rebel attack) can outweigh pro-war effects (a greater capacity to defeat the rebels).

Proposition 1.(iii) establishes two conditions for the selection of the peaceful scenario described in the last paragraph: the foreign actor should be sufficiently uninterested in material costs and benefits (such that the strength of the government is not a big deterrent to intervening), and the government should be sufficiently resourced (such that it can tolerate an attack from the rebels without being seriously harmed).\footnote{If either of these two conditions are not satisfied, Proposition 1.(ii) and Proposition 1.(iii) imply that, in equilibrium, both the government and the rebels will attack (which means that  $(a, a)$ also constitutes a Nash equilibrium). A weak government will attack the rebels (regardless of the risk of foreign intervention) when the government has a limited ability to defend itself from a rebel attack if the government chooses peace.}  To the best of my knowledge, this result is new in the literature. In the next section, I show that it is crucial to understanding the intriguing peace deal between the Colombian government and the FARC.

\section{Case study evidence from Colombia}\label{caseevidence}

In this section, I show how the model proposed in the previous section explains case study evidence from Colombia. I use information from the main actors in the Colombian conflict, as well as from the secondary academic literature. 

\subsection{Background}

The conflict between the Colombian government and the left-wing rebels dates back to the late 1950s.  Its origins have been associated with the founding of the FARC, which was always Colombia's largest and best-equipped rebel group.\footnote{Other smaller rebel groups participated in Colombia's conflict. These include other left-wing insurgents and right-wing paramilitaries. The most important left-wing insurgent other than the FARC is the National Liberation Army (ELN). The main right-wing paramilitary group was the United Self-Defense Forces of Colombia (AUC), which officially demobilized in 2006.} Between 1958 and 2012, the conflict claimed at least 220,000 lives \cite[see][p. 31]{GMH2013}.

From the 1980s until its end, the Colombian conflict was accompanied by several negotiations, including three failed peace talks with the FARC \cite[see][]{Sanchez2001, Chernick2009, Nasi2009}. This section focuses on the most recent (and successful) peace talks, which occurred from 2010 to 2016. These peace talks were preceded by the ``Cagu\'an'' peace process, which occurred from 1998 to 2002, and which  failed miserably.  A relatively weak government, amateurish bargaining teams and spoilers (actors who use violence to undermine peace talks) are some of the explanations given for Cagu\'an's failure \cite[see][]{Kline2007, Nasi2009}. The breakdown of the Cagu\'an peace process led to the election, in 2002, of a hawkish and far-right president, Alvaro Uribe, and started a period of intense war between the government and the FARC. 

\subsubsection{Government empowerment}\label{Governmentempowerment}

During the 2000s and early 2010s, Colombia's government pursued an all-out military effort against the FARC. Due to an increase in defense spending and a significant improvement in military effectiveness, the Colombian government achieved relative success: during this period, the FARC suffered the worst blows in its history. According to the Colombian Ministry of Defense, an average of 40 FARC members were captured or killed annually from 2002 and 2011. In the same period, roughly 17 members deserted the FARC each year, and the number of FARC combatants was halved.\footnote{See \cite{MInDef2009} and \url{https://www.youtube.com/watch?v=vnuqZOwOBoo}. The exact number of FARC members who deserted or were captured or killed has been debated by analysts \citep{Avila2013, RazonPublicaApril2013} and the FARC \cite[Feb. 12]{AnncolFEB122013}, but analysts agree with the government that the number of combatants fell by roughly half during the 2000s.} The government also killed the head of the FARC and several other leaders, in an action characterized by the Colombian president as ``the most devastating blow that this group has suffered in its history.''\footnote{Translation by the author from \url{https://www.youtube.com/watch?v=UI0CJzJLsxU}.} These actions were deeply resented by the FARC.\footnote{See \url{http://www.rebelion.org/noticia.php?id=138858}.}

During this period, the FARC retreated from key regions in the center of the country (i.e. the departments of Cundinamarca, Tolima, and Santander) to border areas with Venezuela and Ecuador (i.e. to the departments of Nari\~{n}o, Cauca, Caquet\'a, Norte de Santander and Arauca) \cite[see][]{IISS2011, Avila2013}. The FARC's decision to move to the periphery of the country was strategic, given ideological similarities between the FARC and the governments of Venezuela and Ecuador, and the porousness of the borders with these two countries \cite[see][p. 257]{Owen2010}. The FARC's decision, combined with a hawkish military strategy by the Colombian government, brought the three countries to the brink of war. I argue that it also raised the likelihood of a peaceful solution.

\subsubsection{On the brink of war}\label{brinkwar}

In March 2008, Colombian security forces crossed into Ecuador to assault an outpost of the FARC. More than two dozen rebels were killed, including a high-ranking leader thought by many to be FARC's second-in-command. The Colombian government also captured computers with documents indicating that Venezuela had been supporting the FARC \cite[see][]{IISS2011}.
 
The assault caused a serious diplomatic incident between Colombia and Ecuador. Ecuador immediately broke off diplomatic relations with Colombia. Venezuela, in solidarity with Ecuador, expelled Colombia's ambassador and other diplomats.\footnote{See  \citet[Mar. 4]{TheNYTimesMARCH42008}} Venezuela and Ecuador also sent troops to the Colombian border, advising that any additional violations of their sovereignty would result in war.\footnote{See \url{https://www.youtube.com/watch?v=Xp7Gs1-tm1w}.} 
The tension reached a peak in July 2010 when, weeks before a change in Colombia's government, the Colombian press secretary provided evidence of a FARC presence in Venezuela to international authorities.\footnote{According to the Colombian press secretary, ``For six years the Colombian government sustained a patient dialogue with the Venezuelan government, on various occasions providing it information on the location of terrorists in that territory. All was unsuccessful with respect to terrorist leaders. We must once again consider taking the matter to international authorities.'' See \citet[July 16]{COLPresJUL162010}.}

\smallskip
In the following days, the Colombian Ambassador to the Organization of American States (OAS) presented photographs, maps, coordinates, and videos proving the presence of illegal armed groups in Venezuelan territory.\footnote{See  \citet[July 22]{OASJUL222010} and \url{https://www.youtube.com/watch?v=J2W0EO27yEQ}.} Venezuela reacted by breaking off diplomatic relations with Colombia, sending more troops to the border and ordering them to be on full alert. Venezuela's then-President Hugo Chavez said: 
\vspace{-0.05cm}
\begin{quote}
To maintain our dignity, we do not have any other option but to sever diplomatic ties with Colombia [...] We will be on alert ---I have ordered the maximum alert along our border [...] Uribe is a threat for peace. He is even able to establish a fake camp in our territory and raid it to start a war.\footnote{Translation by the author from \url{https://www.youtube.com/watch?v=ql_AFMvwg9U}. As \cite{SerbinSerbin2017} say, ``this mobilization of troops was not a minor factor [...] Since 1987 [....] there had been no mobilization of the Venezuelan military at this scale'' \cite[see][p. 241]{SerbinSerbin2017}.} 
\end{quote}
\vspace{-0.05cm}
The Colombian government was aware of the high risk of war, particularly if it violated a neighbor's sovereignty. The Colombian Minister of Defense at the time, Gabriel Silva, said: 
\vspace{-0.05cm}
\begin{quote}
I said privately to President Uribe, ``If you give an authorization, I will bring back to Colombia [all FARC leaders] who are there [in Venezuela]'' [...] He did not authorize. He said it was too risky for the country and for national security [...] I do believe that there was very nearly a war with Venezuela. \cite[][p. 89-91; translation of the author]{Davila2014}.
\end{quote}
\vspace{-0.1cm}

\subsubsection{A moment in history}

In July 2010, Colombia severed diplomatic relations with Venezuela. Two weeks later, a new Colombian president, Juan Manuel Santos, took office. Known for his strategic pragmatism, Santos was closely associated with Uribe's successful military campaigns against the FARC and elected with a mandate to continue Uribe's hard-line policies.\footnote{See \citet[June 12]{SemanaJUNE122010}.}

From his early days in office, Santos combined an extremely aggressive campaign against the FARC with efforts to improve diplomatic relations with Venezuela. Ten days after he was sworn in, diplomatic relations with Venezuela were restored, and approximately one month later, Santos announced the death of the FARC's second-in-command and leader of its strongest fighting division. One year later, the FARC's top leader, who went by the nom de guerre of Alfonso Cano, was killed. The FARC's choice for Cano's replacement, whose nom de guerre was Timochenko, was also influential: Timochenko was known for operating along the border with Venezuela and for having lived there previously.\footnote{See  \citet[Nov. 15]{EltiempoNOV152011} and \citet[Nov. 19]{EltiempoNOV192011}.} In addition, many people had raised concerns about the close ties between Timochenko and important figures in the Venezuelan government.\footnote{Timochenko was called the FARC's ``ambassador'' to Venezuela \citep[see][Mar. 2]{CaracolMarch22010}.} 

In September 2012, almost one year after Timochenko had become the FARC's leader, the Colombian president announced that his government and the FARC had agreed to start a peace process. The announcement surprised analysts and national leaders.\footnote{See \citet[Sept. 3]{SemanaSEP32012} for initial reactions.}  These peace talks were the first open negotiations in a decade. In July 2016, Colombian President Juan Manuel Santos and the FARC signed a historic peace deal.\footnote{See \citet[Sept. 26]{TheNYTimesSEPT262016}.} In October 2016, Juan Manuel Santos was awarded the Nobel Peace Prize for his efforts to end the Colombian civil war.\footnote{See \citet[Oct. 7]{TheNYTimesOCT72016}.}

\subsection{Analysis}

Why did Colombia and the FARC decide to pursue peace? I suggest that the risk of externalization of the Colombian conflict to Venezuela, the politicization of Venezuela's foreign policy, and the military strength of the Colombian government are at the root of the peace deal, creating what some literature on conflict has called a ``ripe for resolution'' situation \cite[e.g. see][]{Zartman2000}.

The FARC's motivations for peace seem clear: it suffered significant setbacks between 2002 and 2011, and its leaders seemed convinced that they had no chance of defeating their enemy. Talking about the relevance of the setbacks, the FARC's leader, Timochenko, said in 2012: ``I can't deny we've received serious blows ---and extremely painful ones.  The deaths of four members of the National Secretariat can't be minimized [...] it's obvious that today's conditions are not the same as a decade ago.''\footnote{\citet[Sept. 19]{LozanowebsiteSEP192013}, translation of the author.} Asked why the FARC decided to negotiate with Santos, Timochenko responded that the costs of continuing the conflict would have been very high:\footnote{The FARC's strategy may have always been respond to peace with peace. According to the FARC's leader, the FARC ``negotiate because a political solution has always been our objective, and also that of the people's movement'' \cite[][Sept. 19, translation of the author]{LozanowebsiteSEP192013}. This awareness was also recognized by FARC's representative of the peace delegation, Pablo Catatumbo:  ``We are ready to start preparing the way that will lead us towards the expression of our regret for what has happened ... No doubt there has also been harshness and pain caused from our side.'' \cite[][translation of the author]{FARCaug202013}.} 
\vspace{-0.06cm}
 \begin{quote}
Whatever may come, persistent conflict will entail many more deaths and great destruction, more sorrow and tears, more poverty and misery for some and greater wealth for others. Imagine the lives that could have been saved over the past 10 years. That's why we seek negotiations, a solution without blood, and an understanding through political routes. \cite[][Sept. 19; translation of the author]{LozanowebsiteSEP192013}.
\end{quote}
\vspace{-0.0cm}
The Colombian government's motivations for negotiating with the FARC are less apparent. Analysts and the Colombian government have highlighted  the increase in the power of Colombian forces relative to the FARC.\footnote{For instance, Colombian president Santos said in 2012, ``If we can talk about peace now [...] it is because of the effectiveness of our armed forces."  \cite[See][Oct. 25, translation of the author]{COLPresOCT252012}.} But if the Colombian government was winning the war, why wouldn't it continue fighting the FARC for a few more years, as advocated for by former president Uribe?\footnote{See \citet[Oct. 28]{CaracolOCT282012}.} War is costly and unpredictable, so rational agents should have incentives to reach peaceful settlements that all would prefer to war. However, from the point of view of the majority of the Colombian population, media, and important leaders, just before the peace talks were made public, the FARC were close to being defeated. Thus, the government's cost of war could be perceived as being very low. In addition, at the time, a peace process seemed riskier, and the FARC was widely seen as untrustworthy.\footnote{In the speech announcing the opening of peace talks, Santos said, ``There comes a moment in history when you have to take risks to arrive at a solution [...] This is one of those moments." (See \citealp[Sept. 4]{COLPresSEPT42012}, translation of the author).} I argue that the main reasons the government chose peace was the high risk of an international conflict with Venezuela,  the extreme politicization of Venezuela's foreign policy, and the Colombian government's significant military strength. According to the theory proposed in Section \ref{model}, these conditions are sufficient for a peaceful equilibrium. 

Were these conditions satisfied just before the start of the peace talks between the Colombian government and the FARC? I already argued (in  subsections \ref{Governmentempowerment} and \ref{brinkwar}) that two of these conditions were likely  satisfied. First, I showed that just before the start of the peace talks, the government was as strong as ever. A military victory by the FARC  (or any other non-state armed group) was seen as so unlikely that the discussion was entirely focused on when --- not if --- the government would defeat the guerrilla groups. Second, I showed that just before Santos took power, the risk of an interstate conflict between Colombia and Venezuela was extremely high. This risk was directly related to the presence of the FARC in Venezuelan territory, the very likely possibility that Colombia might violate Venezuelan sovereignty to pursue the rebels, and Venezuela's determination to respond if this happened. 

The third condition --- the extreme politicization of Venezuela's foreign policy --- was also met.  According to \cite{SerbinSerbin2017}, Venezuela's oil-based foreign activism was amplified under the presidency of Hugo Chavez, and, importantly, included: 
%
 \begin{quote}
 ``the unbridled involvement of the president in foreign affairs and its extreme politicization [...] [and this trait was expressed in] the role assigned to the armed forces  in dealing with a potential asymmetric conflict with the United States, in developing links and military exchanges with countries in the region and beyond, and in establishing ties with irregular forces such as the guerrilla movements'' \cite[][p. 239]{SerbinSerbin2017}
 \end{quote}
 %
The conflict with Colombia just before the start of the peace talks seems to be a perfect example of this activism: Colombian had Latin America's most pro-American president,  the Colombian armed forces were perceived by Chavez as being commanded  by the United States,\footnote{Motivated by Colombia's plans to allow the US to use seven of its military bases, Chavez said in 2009: ``The Yankees are starting to command the Colombian armed forces. They are the ones who are in charge'' \cite[][Agu. 10]{FTAUGUST102009}} and the FARC and Chavez were not only ideologically very close (i.e. both were left-wing ad anti-American), but shared an explicit Pan-American ``Bolivarian'' discourse.\footnote{See \citet[Jan. 19]{SemanaJAN192008}.} 

\section{Conclusion}\label{conclusion}

In this paper, I develop a simple model of conflict externalization and provide new case study evidence from Colombia. The first main contribution of the paper is to show that the risk of externalization of a domestic conflict increases the likelihood of peace, but that this only happens if the domestic government is sufficiently powerful, if the risk of an external intervention is sufficiently high, and if the foreign party that may intervene is sufficiently uninterested in material costs and benefits.

In the second part of the paper, I use the model to examine the Colombian conflict.  I focus on peace talks that occurred between 2010 and 2016 between the Colombian government and the FARC. The second main contribution of the paper is to show how the risk of externalization of the conflict to Venezuela played a crucial role in the success of these talks. 

Although the theory is inspired by the Colombian conflict, its application is not limited to this case. The model can be applied to any internal conflict in which governments, undertaking cross-border counterinsurgency actions, initiate military actions against neighboring states. While other explanations exist to explain how Colombian peace talks evolved, such as less ideological extremism and less militarism from both parties, the evidence shows that the possibility of externalization should be considered in any examination of the issue.

\newpage

\section*{Appendix}

\small

\begin{proof}[Proof of Proposition 1] I solve the game by backward induction. Having examined $F$'s  decision about whether to intervene in the main text, I must now examine the other groups' optimal decisions about whether to attack each other. I start by showing that, under Assumption 1, peace (i.e., $(p, p)$) always constitutes a Nash equilibrium of the game. Then, I show that under Assumptions 2 and 3, there is a threshold value for $\phi$ that determines whether war (i.e., $(a, a)$) is also a Nash equilibrium of the game. I conclude by showing that neither $(a, p)$ nor $(p, a)$ can constitute a Nash equilibrium. 

\medskip

First, define $\pi(p,p)-\pi(a,p)=Z(G)-(1-\phi)(1-W(G)) Z(G+L)+C$, where $ \pi(p,p)$ and $\pi(a,p)$ are from Table \ref{payoffs}, and where I have used the fact that $\Phi=\phi +(1-\phi)W(G)$. Now note that if $\pi(p,p)>\pi(a,p)$, then when the rebels seek peace, the government's best response is also to seek peace. In the last expression, note that $\pi(p,p)>\pi(a,p)$ if and only if 
\begin{equation}
\label{ebiggerzero}
C>(1-\phi)(1-W(G)) Z(G+L)-Z(G).
\end{equation}
Now note that since $Z(0)=0$, $Z(x)=1$ for all $x\geq \overline{G}$, and $Z'>0$ and $Z''\leq 0$ for all $x\in(0,\overline{G})$, then Assumption 1  implies that $C> Z(x +L)-Z(x)$  for $x\geq 0$. In particular, Assumption 1  implies  that $C> Z(G +L)-Z(G)$ for $G\in(L,\overline{G})$. Thus, since $(1-\phi)(1-W(G))\leq 1$ for all  $G\in(L,\overline{G})$, we have that (\ref{ebiggerzero}) holds for all $G\in(L,\overline{G})$. 

Now define $\rho(p,p)-\rho(p,a)=-Z(G)+Z(G-L)+C$, where $\rho(p,p)$ and $\rho(p,a)$ are from Table \ref{payoffs}, and where I have used the fact that $\Phi=\phi +(1-\phi)W(G)$. Note that if $\rho(p,p)>\rho(p,a)$, the rebels' best response is to seek peace, given that the government also seeks peace. In the last expression, note that $\rho(p,p)>\rho(p,a)$ if and only if 
\begin{equation}
\label{fbiggerzero}
C>Z(G)-Z(G-L).
\end{equation}
Since $C> Z(x +L)-Z(x)$  for all $x\geq 0$ (which follows from Assumption 1, as previously shown), then choosing $x=G-L$ for $G\in(L,\overline{G})$,  we have that (\ref{fbiggerzero}) holds for all $G\in(L,\overline{G})$.   

We therefore have that peace (i.e., $(p, p)$)  constitutes a Nash equilibrium of the game. This proves Proposition 1.(i). 

\medskip

Now I establish the conditions under which war (i.e., $(a, a)$) is also a Nash equilibrium of the game. To do this, first define $D\equiv \pi(p,a)-\pi(a,a)$, and note that if $D>0$, then the government's best response to a rebel attack is to choose peace, and if $D<0$, the government's best response is to attack.  From Table \ref{payoffs}, note that $D$ is equal to
\begin{equation}
\label{dbiggerzero}
D=Z(G-L)-(1-\phi)(1 -W(G)) Z(G)
\end{equation}
where I have used the fact that $\Phi=\phi +(1-\phi)W(G)$. Differentiating (\ref{dbiggerzero}) with respect to $G$, we get
\begin{equation}
\label{dDdg1}
\frac{d D}{dG}=Z'(G-L) - (1 -\phi)(1-W(G))Z'(G)+(1 -\phi)W'(G)Z(G)
\end{equation}
which, rearranging the terms, is equivalent to
\begin{equation}
\label{dDdg2}
\frac{d D}{dG}=[Z'(G-L) - (1 -\phi)Z'(G)]+(1 -\phi)W'(G)Z(G)\Big[1+\frac{W(G)Z'(G)}{W'(G)Z(G)}\Big].
\end{equation}
In (\ref{dDdg2}), note that $Z'(G-L) - (1 -\phi)Z'(G)\geq 0$ since $Z''(x)\leq 0$ for all $x\in(L,\overline{G})$. Thus, given any $\phi \in [0,1]$, a sufficient condition for $\frac{d D}{dG}>0$ is 
\begin{equation}
\label{ass2condproof1}
1+\frac{W(G)Z'(G)}{W'(G)Z(G)}<0
\end{equation}
given that $W'<0$ for all $x\in(0,A)$ where $A>\overline{G}$. I will now show that Assumption 2 implies that (\ref{ass2condproof1}) holds for all $G\in(L,\overline{G})$. To see this, note that
\vspace{-0.0cm}
\begin{equation}
\label{ass2condproof2}
1+\frac{W(G)}{Z(G)}\frac{Z'(G)}{W'(G)}\leq 1+\frac{W(G)}{Z(G)}k
\end{equation}
\vspace{-0.2cm}

where  $k\equiv sup\{\frac{Z'(G)}{W'(G)}|G\in(L,\overline{G})\}$. In addition, note that since $W'(x)< 0$  and $Z'(x)>0$ for all $x\in(0,\overline{G})$, we have that $k\leq 0$, and that 
\begin{equation}
\label{ass2condproof3}
1+\frac{W(G)}{Z(G)}k\leq 1+W(G)k\leq 1+W(\overline{G})k.
\end{equation}
Combining (\ref{ass2condproof2}) and (\ref{ass2condproof3}), we have that 
\vspace{-0.1cm}
\begin{equation}
\label{ass2condproof4}
1+\frac{W(G)}{Z(G)}\frac{Z'(G)}{W'(G)}\leq 1+W(\overline{G})k.
\end{equation}
\vspace{-0.4cm}

So, if $1+W(\overline{G})k<0$ (as stated in Assumption 2), then (\ref{ass2condproof1}) holds.  And if (\ref{ass2condproof1}) holds, we have that (\ref{dDdg2}) is greater than zero given any $\phi \in [0,1]$ and for all $G\in(L,\overline{G})$. 

\medskip

Next, note that $\frac{d D}{dG}>0$ and $L<\overline{G}$ implies that 
\begin{equation}
\label{ }
D\in\big(-(1-\phi)(1-W(L))Z(L),Z(\overline{G}-L)-(1-\phi)(1-W(\overline{G}))Z(\overline{G})\big) 
\end{equation}
for any $G\in(L,\overline{G})$.  Since  $-(1-\phi)(1-W(L))Z(L)<0$ when $\phi<1$, then, for this case, the government's best response to a rebel attack is always to counterattack if $Z(\overline{G}-L)-(1-\phi)(1-W(\overline{G}))Z(\overline{G}) < 0$, or, equivalently, if
\vspace{-0.0cm}
\begin{equation}
\label{overlinephi0}
Z(\overline{G}-L)-(1-\phi)(1-W(\overline{G}))< 0
\end{equation}
where I have used the fact that  $Z(\overline{G})=1$. To identify the conditions under which (\ref{overlinephi0}) holds,  define
\vspace{-0.1cm}
\begin{equation}
\label{overlinephi1}
\overline{\phi}\equiv 1- \frac{Z(\overline{G}-L)}{(1-W(\overline{G}))}
\end{equation}
\vspace{-0.2cm}

and note that under Assumption 3, $\overline{\phi}>0$. From (\ref{overlinephi1}), note that when  $\phi<\overline{\phi}$, (\ref{overlinephi0}) holds. This means that when $\phi<\overline{\phi}$, the government's best response to a rebel attack is to also attack, regardless of the value of $G$. And since the rebel's best response to a government attack is clearly to attack, then whenever $\phi \leq \overline{\phi}$ and  $\phi<1$, $(a, a)$ there is a Nash equilibrium for all $G\in(L,\overline{G})$. This proves Proposition 1.(ii). 

\medskip

From (\ref{overlinephi1}), note that $\overline{\phi}<1$, so there are values of $\phi$ such that $ 1 > \phi > \overline{\phi}$. For these values, note that (\ref{overlinephi0}) does not hold. Combining this observation with the fact that $\frac{d D}{dG}>0$  for all $G\in(L,\overline{G})$ (and  given any $\phi \in [0,1]$), we obtain that whenever $1 > \phi > \overline{\phi}$, there exists $\hat{G}(\phi)\in(L,\overline{G})$, defined implicitly by 
\vspace{-0.15cm}
\begin{equation}
\label{defGhat}
Z(\hat{G}-L)=(1-\phi)(1 -W(\hat{G}))Z(\hat{G})
\end{equation}
such that $(a, a)$ constitutes a Nash equilibrium if and only if $G \leq \hat{G}(\phi)$. To see that $\hat{G}'(\phi) < 0$, differentiate implicitly (\ref{defGhat}) with respect to $\phi$ to get $\hat{G}'(\frac{d D}{d G}|_{G=\hat{G}})+(1 -W(\hat{G})) Z(\hat{G})=0$; then note that the fact that for any $\phi \in [0,1]$, $\frac{d D}{dG}>0$  for $G\in(L,\overline{G})$, and that $(1 -W(\hat{G})) Z(\hat{G})>0$, imply that $\hat{G}'<0$.  This proves Proposition 1.(iii). 

\bigskip

Finally, recall from the analysis above that the best response of both the rebels and government is to seek peace when their opponent also seeks peace. This implies that neither $(a, p)$ nor $(p, a)$ can constitute a Nash equilibrium (i.e., Proposition 1.(v)). This, combined with the fact that when $\phi = 1$, $D=Z(G-L)>0$ for all $G\in(L,\overline{G})$ proves Proposition 1.(iv). 

\black

\end{proof}

\newpage
\def\url#1{}
\bibliographystyle{aer}
\bibliography{bibexternalization}

\end{document}